\documentclass[blue]{ludusofficial}  % Theme: blue, red, green, purple, orange, cyan
\usepackage{amsmath}
\usepackage{dblfloatfix}
\usepackage{placeins} 
\title{Longitudinal Boundary Sharpness Coefficient Slopes Predict Time to Alzheimer's Disease Conversion in Mild Cognitive Impairment:\\[8pt]A Survival Analysis Using the ADNI Cohort}

\shorttitle{BSC Slopes Predict MCI-to-AD Conversion}  % Short title for page headers
\shortauthor{Cherukuri}  % Short author names for page headers

\author{%
    \textbf{Ishaan Cherukuri}\textsuperscript{1}\textsuperscript{*}\\[0.2cm]
    \small\textsuperscript{1}Independent Researcher, USA\\
    \small Corresponding: ishaan.cherukuri@gmail.com\\[0.4cm]
    \small\textsuperscript{*}Data used in preparation of this article were obtained from the Alzheimer's Disease Neuroimaging Initiative (ADNI) database (adni.loni.usc.edu). As such, the investigators within the ADNI contributed to the design and implementation of ADNI and/or provided data but did not participate in the analysis or writing of this report. A complete listing of ADNI investigators can be found at: \href{http://adni.loni.usc.edu/wp-content/uploads/how_to_apply/ADNI_Acknowledgement_List.pdf}{adni.loni.usc.edu/how\_to\_apply/ADNI\_Acknowledgement\_List.pdf}
}

\begin{document}

%% Abstract and keywords must be defined BEFORE \maketitle
\begin{abstract}
Predicting whether someone with mild cognitive impairment (MCI) will progress to Alzheimer's disease (AD) is crucial in the early stages of neurodegeneration. This uncertainty limits enrollment in clinical trials and delays urgent treatment. The Boundary Sharpness Coefficient (BSC)\cite{olafson2021bsc} measures how well-defined the gray-white matter boundary looks on structural MRI. This study measures how BSC changes over time, namely, how fast the boundary degrades each year works much better than looking at a single baseline scan for predicting MCI-to-AD conversion. This study analyzed 1,824 T1-weighted MRI scans from 450 ADNI subjects (95 converters, 355 stable; mean follow-up: 4.84 years). BSC voxel-wise maps were computed using tissue segmentation at the gray-white matter cortical ribbon. Previous studies have used CNN and RNN models that reached 96.0\% accuracy for AD classification and 84.2\% for MCI conversion~\cite{hafeez2024deep}, but those approaches disregard specific regions within the brain. This study focused specifically on the gray-white matter interface. The approach uses temporal slope features capturing boundary degradation rates, feeding them into Random Survival Forest~\cite{article}, a non-parametric ensemble method for right-censored survival data. The Random Survival Forest trained on BSC slopes achieved a test C-index of 0.63, a \textbf{163\% improvement} over baseline parametric models (test C-index: 0.24). Structural MRI costs a fraction of PET imaging (\$800--\$1,500 vs. \$5,000--\$7,000) and does not require CSF collection. These temporal biomarkers could help with patient-centered safety screening as well as risk assessment.
\end{abstract}


%% Maketitle creates full-width header with abstract and keywords, then starts two-column layout
\maketitle

%% From here, content is in two columns
\section{Introduction}

Alzheimer's disease (AD) already affects more than 55 million people worldwide, and projections suggest that figure could rise to 131 million by 2050~\cite{hafeez2024deep}. It remains the leading cause of dementia, accounting for roughly 60--80\% of cases, and the economic burden is staggering, now exceeding \$1 trillion globally each year. Numbers like these can start to feel abstract. In practice, though, AD is a disease that gradually strips away memory, judgment, daily independence, and eventually the basic structure of ordinary life. Biologically, that decline is tied to amyloid-$\beta$ plaques, tau tangles, synaptic dysfunction, and progressive brain atrophy.

Mild Cognitive Impairment (MCI) sits in a much more uncertain space. It is not normal aging, but it is not yet dementia either. People with MCI often notice memory or thinking problems that matter, though those changes are still not severe enough to fully disrupt independent living. The difficulty is that MCI does not lead everyone down the same path. Annual conversion rates to dementia, especially AD, are often estimated around 10--15\%~\cite{li2021predicting}, yet that average hides a lot. Some individuals remain stable for years. Some even appear to improve. Others decline quickly. That uncertainty is precisely why early risk stratification matters. Treatments and interventions are most likely to help before too much irreversible damage has accumulated.

The urgency has only increased with the arrival of anti-amyloid therapies such as lecanemab~\cite{vanDyck2022lecanemab} and donanemab~\cite{sims2023donanemab}. These drugs are generally aimed at earlier stages of disease, which means clinicians and researchers need better ways to identify who is likely to progress, and when. Existing tools help, but each comes with tradeoffs. Amyloid PET can detect pathology more directly, yet it is expensive, often costing \$5,000--\$7,000 per scan, and it is not widely accessible. Cerebrospinal fluid testing provides valuable molecular information, but lumbar puncture is invasive and not always acceptable to patients. Blood-based biomarkers are increasingly promising and are beginning to enter practice, although interpretation is still not entirely straightforward because false positives and false negatives remain possible. By contrast, structural MRI is already part of standard workups for cognitive complaints, costs substantially less at roughly \$800--\$1,500, and does not require additional procedures. The catch is that conventional MRI markers, especially when measured at only one time point, have often been disappointing for prediction~\cite{baytas2024predicting}.

\subsection{Prior Work}

A wide range of models has been proposed to predict progression from MCI to AD. Li et al.~\cite{li2021predicting}, for example, reported an Area Under the Curve (AUC) of 83\% by combining genetic data, gene expression, and neuroimaging. That is impressive, but it also points to a practical problem. A model can perform well in theory while remaining difficult to deploy in ordinary clinics. Multimodal pipelines often demand blood collection, specialized assays, and higher per-patient cost, sometimes in the \$2,000--\$5,000 range.

Other groups have taken a different route by mining electronic health records. Some of those studies report accuracies around 72\% even several years before diagnosis~\cite{north2025breakthroughs}. That work is useful, especially at scale, but it mostly captures downstream clinical traces such as medication history, diagnoses, and laboratory patterns. It does not directly measure what is happening in brain tissue itself.

Single-timepoint neuroimaging has similar limitations. Hafeez et al.~\cite{hafeez2024deep} showed that deep learning models could classify AD with 96\% accuracy from imaging snapshots, yet performance dropped to 84\% when the task shifted to predicting MCI conversion. That gap is telling. Identifying established disease is not the same as forecasting future decline. A baseline scan may reveal what the brain looks like at one moment, but it does not necessarily show where that brain is heading. The central idea of the present study is therefore fairly simple, though maybe more consequential than it first sounds: \textit{how fast a biomarker changes may carry more predictive value than the biomarker's starting level alone}.

\subsection{The Boundary Sharpness Coefficient}

The Boundary Sharpness Coefficient (BSC) measures how sharply gray matter and white matter are separated on structural MRI. That may sound technical, but the intuition is not too complicated. In a healthy brain, the transition between the cortical gray matter and the underlying white matter is usually fairly crisp on a T1-weighted scan. Gray matter tends to appear darker, while white matter is brighter because of its myelin-rich composition. When disease-related processes disrupt that organization, the interface can become less distinct.

BSC is therefore different from more familiar measures such as regional volume or cortical thickness. Those metrics tell us how much tissue remains. BSC, instead, captures a boundary property. It is not a direct microscope-like readout of microstructure, and it would be overstating things to claim otherwise. Still, it may reflect biologically meaningful changes at the gray-white matter interface before more obvious atrophy becomes visible. In that sense, it occupies an interesting middle ground between gross anatomy and finer tissue organization.

Several pathological processes in AD could plausibly blur this boundary. Amyloid and tau pathology disrupt neuronal integrity. Synaptic loss changes cortical organization. Inflammation alters tissue composition. Myelin degeneration affects white matter signal properties. None of those processes maps one-to-one onto BSC, but together they provide a reasonable biological basis for why boundary sharpness might deteriorate as disease advances.

BSC computation in this study followed a multi-step pipeline (Figure~\ref{fig:pipeline}). First, Atropos k-means clustering~\cite{avants2011atropos} assigned each voxel a probability of belonging to gray matter, white matter, or cerebrospinal fluid. Next, boundary voxels were defined as those where gray matter probability fell between 0.4 and 0.6, which captures a transition band rather than forcing the analysis onto an unstable single-voxel contour. Gradient filters then measured how rapidly MRI intensity changed across that boundary, and the gradient was projected along the direction perpendicular to the interface. For each boundary voxel $\mathbf{x}$, the directional BSC was computed as follows:

\begin{figure*}[t!]
\centering
\includegraphics[width=0.95\textwidth]{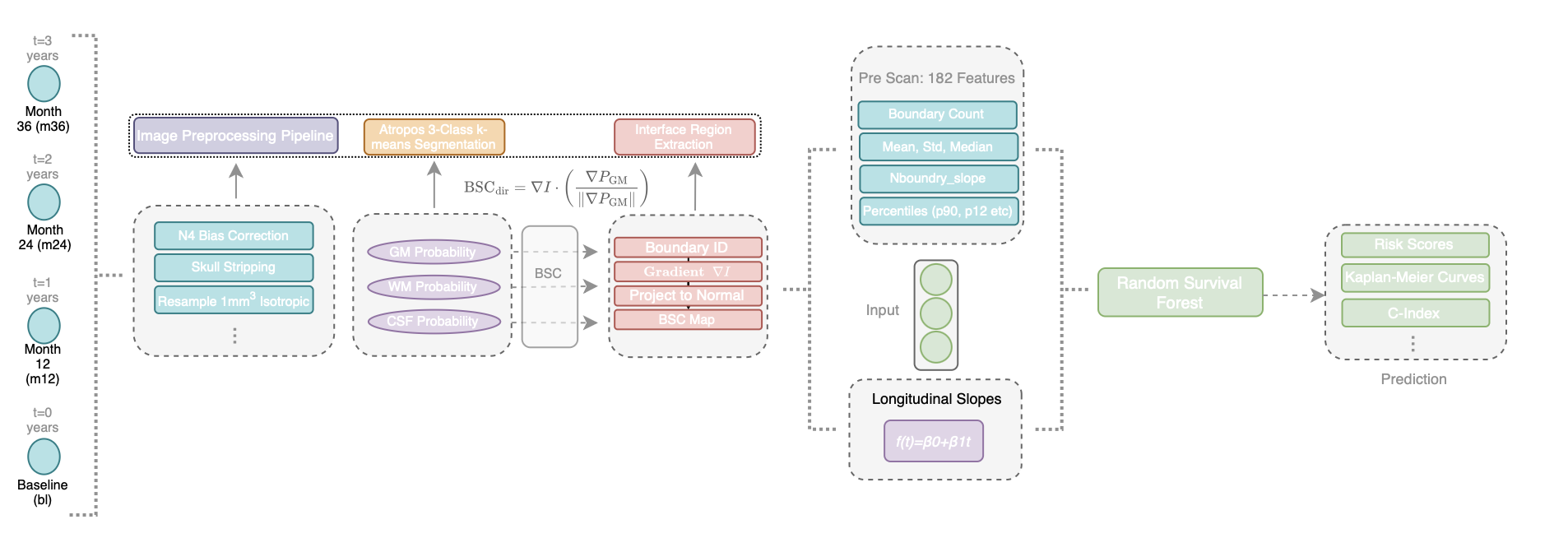}
\caption{Complete BSC slope-based MCI-to-AD conversion prediction pipeline. Longitudinal T1-weighted MRI scans from four time points (baseline, m12, m24, m36) undergo preprocessing (N4 bias correction, skull stripping, resampling), Atropos segmentation to derive GM/WM probability maps, and BSC computation at the gray-white matter boundary. Per-scan features (35 total) are extracted from each time point, then organized into per-subject longitudinal sequences. Linear regression across time points yields 182 slope-based features per subject capturing annual rates of boundary degradation. The top 20 most variable slopes are selected and input to a Random Survival Forest model, which outputs individual risk scores, Kaplan-Meier survival curves stratified by risk group, and performance metrics (C-index: 0.63).}
\label{fig:pipeline}
\end{figure*}

\begin{equation}
    \text{BSC}_{\text{dir}}(\mathbf{x}) = \nabla I(\mathbf{x}) \cdot \frac{\nabla P_{GM}(\mathbf{x})}{\|\nabla P_{GM}(\mathbf{x})\|}
\end{equation}

Here, $I(\mathbf{x})$ denotes MRI intensity at voxel $\mathbf{x}$ and $P_{GM}$ is the gray matter probability map. The term $\nabla I$ measures how quickly intensity changes in space, while the dot product isolates the component aligned with the boundary normal. Positive values indicate the expected increase in intensity when moving from gray matter toward white matter. Negative values usually reflect artifacts or local reversals.

The final output is a three-dimensional map in which each boundary voxel is assigned a sharpness value. Those voxel-level measurements are then summarized using means, medians, standard deviations, and percentiles so they can be used in downstream statistical models.

\subsection{From Static Measures to Temporal Slopes}

The weakness of single-timepoint biomarkers is not merely that people have different head sizes or different baseline anatomy. Those issues can often be normalized. The deeper problem is that a cross-sectional scan mixes together many influences that have accumulated across a lifetime. Two people can show the same hippocampal volume, for instance, while meaning very different things biologically. One may simply have had a smaller hippocampus for years. The other may be in the middle of active neurodegeneration. A single snapshot cannot really distinguish those scenarios.

Cognitive reserve complicates the picture even more. Some individuals tolerate substantial pathology before symptoms become obvious. Education, occupational complexity, social engagement, and lifelong learning may all contribute to that resilience. So a person can look relatively intact on standard cognitive tests while disease processes are already advancing beneath the surface.

That is why rates of change are so appealing. A declining trajectory can reveal ongoing pathology in a way that baseline values often cannot. The working hypothesis here is that \textbf{longitudinal BSC slopes}, meaning annualized rates of change in boundary sharpness features, capture progression dynamics that are largely invisible in a baseline scan. Someone whose BSC is deteriorating quickly may face a very different future from someone who starts at a similar level but remains stable over time. More broadly, this aligns with a shift in neurodegeneration research away from static snapshots and toward disease trajectories.

\subsection{Study Objectives}

This study was guided by four main aims:

\begin{enumerate}
    \item Compute longitudinal BSC slopes for 450 ADNI subjects with at least 4 MRI scans each
    \item Train Random Survival Forest (RSF) models to predict time-to-AD-conversion using those slopes
    \item Compare performance against baseline BSC features and parametric AFT models
    \item Identify which BSC slope features are most informative for prediction
\end{enumerate}

\section{Materials and Methods}

\subsection{Dataset and Study Population}

All data were obtained from the Alzheimer's Disease Neuroimaging Initiative (ADNI) database (\texttt{adni.loni.usc.edu})~\cite{weiner2010adni,petersen2010adni}. ADNI began in 2003 as a large public-private effort led by Michael W. Weiner, MD, with the goal of determining whether serial MRI, PET, fluid biomarkers, and cognitive testing could track the progression of MCI and early AD. Over time, ADNI has become one of the most widely used open neuroimaging resources in the field. Its strengths are fairly clear: standardized protocols, repeated follow-up, and broad data sharing. At the same time, like many research cohorts, it is not a perfect mirror of routine clinical populations, which matters when thinking about generalizability.

\subsubsection{Inclusion Criteria}

Subjects were included if they met four criteria: baseline diagnosis of MCI, at least 4 longitudinal T1-weighted MRI scans acquired on different dates, chronologically ordered scans with valid visit labels (bl, m12, m24, m36, and beyond), and complete clinical metadata including diagnosis at each visit. After correcting the imaging manifest for visit order and label inconsistencies, 456 subjects met these criteria.

Among them, \textbf{101 converters} progressed from MCI to AD during follow-up, while \textbf{355 stable subjects} remained classified as MCI or reverted to cognitively normal status without ever receiving an AD diagnosis code of 3.0. Six subjects were excluded because they were already labeled as AD at baseline, which yields a time-to-conversion of zero years. Those cases likely reflect labeling problems or pre-baseline progression and are also incompatible with parametric accelerated failure time models that require positive event times. The final cohort therefore included \textbf{450 subjects}.

\subsection{MRI Acquisition and Preprocessing}

ADNI participants underwent T1-weighted structural MRI on 1.5T or 3T scanners using standardized acquisition protocols~\cite{jack2008adni_mri}. Scans were acquired with 3D magnetization-prepared rapid gradient-echo (MPRAGE) sequences across sites, with typical parameters of TR = 2300 ms, TE = 2.98 ms, TI = 900 ms, flip angle = 9$^\circ$, and isotropic voxel size of 1$\times$1$\times$1 mm$^3$. These settings are widely used because they provide good gray-white matter contrast without excessive scan time.

The BSC pipeline included several preprocessing steps. First, \textbf{N4 bias correction}~\cite{tustison2010n4} was applied to reduce low-frequency intensity non-uniformity caused by magnetic field inhomogeneity. Without that correction, gradual brightness shifts across the image could be mistaken for biologically meaningful boundary changes. N4ITK was run using default convergence settings (threshold: 0.001) and multi-resolution optimization across four levels ([50,50,50,50] iterations).

Next, \textbf{skull stripping} removed non-brain tissues such as skull, scalp, and orbital structures, which otherwise can interfere with segmentation and boundary detection. An Otsu-thresholding approach with morphological cleanup, including largest connected component selection and hole filling, was used. More advanced deep learning skull-stripping methods exist, such as SynthStrip and HD-BET~\cite{hoopes2022synthstrip}, and in some contexts they may perform better. However, the present pipeline prioritized speed, reproducibility, and minimal dependence on external training data.

All images were then \textbf{resampled to 1 mm$^3$ isotropic resolution} using linear interpolation. This step standardized spatial resolution across acquisitions and made voxel-wise comparisons more consistent. Linear interpolation was selected because it avoids some of the ringing artifacts that can appear with higher-order interpolation methods, which is relevant when gradients are later computed.

\textbf{Tissue segmentation} was performed using Atropos $k$-means clustering~\cite{avants2011atropos} with three classes corresponding to gray matter (GM), white matter (WM), and cerebrospinal fluid (CSF). The choice of $k$-means over alternatives such as Gaussian mixture models or deep learning segmentation was mainly pragmatic. It is fast, deterministic enough for reproducible workflows, and generally adequate for T1-weighted tissue separation in large cohorts. Each voxel was assigned class probabilities summing to 1.0, with $P_{GM}(\mathbf{x})$ denoting the gray matter probability at voxel $\mathbf{x}$.

\textbf{Boundary identification} was based on voxels satisfying $0.4 \leq P_{GM} \leq 0.6$, thereby defining a transition band centered around the gray-white matter interface~\cite{olafson2021bsc}. A single exact contour might seem cleaner in theory, but in practice it can be unstable and highly sensitive to noise. The band-based definition is somewhat less elegant, perhaps, yet more robust.

\textbf{Gradient computation} used 3D Gaussian derivative filters with $\sigma = 1.0$ mm to estimate spatial derivatives of both MRI intensity $I(\mathbf{x})$ and gray matter probability $P_{GM}(\mathbf{x})$. The smoothing helps suppress high-frequency noise while preserving edge-related information. This produces gradient vectors $\nabla I$ and $\nabla P_{GM}$ at each voxel. Finally, \textbf{directional projection} extracted the component of the intensity gradient aligned with the boundary normal by taking the dot product with the normalized gray matter probability gradient. In effect, this isolates intensity change across the boundary rather than along it.

\subsection{Longitudinal Slope Computation}

For each scan, 35 BSC summary features were extracted across several categories (Table~\ref{tab:bsc-features}). These included boundary count, directional and magnitude-based summary statistics, percentile measures, and spatial bin summaries. To model longitudinal change, each feature was regressed against time within subject.

\begin{table}[h!]
\centering
\caption{BSC feature categories extracted per scan}
\label{tab:bsc-features}
\begin{tabular}{lc}
\hline
\textbf{Feature Category} & \textbf{Count} \\
\hline
Boundary count (Nboundary) & 1 \\
Directional statistics (mean, std, median) & 3 \\
Directional percentiles (p10, p25, p50, p75, p90) & 5 \\
Magnitude statistics (mean, std, median) & 3 \\
Magnitude percentiles (p10, p25, p50, p75, p90) & 5 \\
Directional spatial bins (8 bins) & 8 \\
Magnitude spatial bins (8 bins) & 8 \\
\hline
\textbf{Total per scan} & \textbf{35} \\
\hline
\end{tabular}
\end{table}

For subject $s$ with scans acquired at times $t_1, t_2, t_3, t_4, \ldots$, longitudinal slopes for each feature $f$ were estimated using linear regression:

\begin{equation}
    f_s(t) = \beta_0 + \beta_1 \cdot t + \epsilon
\end{equation}

where $\beta_1$ represents the annualized rate of change. Actual acquisition dates, rather than nominal visit codes, were used as the time variable so that irregular spacing between scans was captured accurately.

For each of the 35 base features, four primary longitudinal descriptors were derived: baseline value ($f_{\text{baseline}}$), final value ($f_{\text{final}}$), annual slope ($\beta_1$), and goodness-of-fit ($R^2$). Together these yielded \textbf{140 primary derived features} (4 $\times$ 35). In addition, \textbf{42 interaction and ratio features} were calculated, including absolute change ($f_{\text{final}} - f_{\text{baseline}}$), percent change, confidence-weighted slopes ($\beta_1 \times R^2$), percentile ratios such as p90/p10, and spatial asymmetry indices contrasting left and right hemisphere bins. In total, \textbf{182 slope-based features per subject} were generated for survival modeling.

\subsection{Feature Selection and Preprocessing}

The raw slope features showed extreme heterogeneity in variance. In particular, Nboundary\_slope, which reflects the rate of change in boundary voxel count, varied on a scale far larger than most other features. Left untreated, that imbalance risked allowing a single feature family to dominate downstream model behavior.

To address this, a feature selection and preprocessing pipeline was designed with robustness in mind. Selection was carried out on the training data only to avoid information leakage. First, raw slope values were transformed using a signed logarithm, $\text{sign}(x) \cdot \log(1 + |x|)$, which compresses extreme values while preserving the direction of change. Next, outliers were clipped by winsorization~\cite{winsorization} at the 1st and 99th percentiles, with those limits fit on the training set and then applied to the test set. Variance ranking was then computed on these robustly transformed features.

Because Nboundary-derived measures still tended to overwhelm the ranking, a \textbf{0.10 penalty factor} was applied to boundary count features during the selection step. This penalty altered the ranking scores only, not the actual feature values used for modeling. The goal was not to remove boundary count information, which may be biologically meaningful, but to prevent the model from reducing the analysis to a near-univariate problem.

After ranking, the top 20 features were retained. These selected features were passed through the same signed-log and winsorization steps, followed by MinMax scaling to the range [0, 1]. All transformation parameters were estimated on the training set alone and then transferred to the test set. After preprocessing, feature variances in the training data ranged from 0.018 to 0.142, a ratio of 7.9:1, compared with an original raw variance ratio above 300{,}000:1. That reduction substantially improved numerical balance across inputs while keeping directional interpretation intact.

\subsection{Survival Data Preparation}

Survival outcomes were defined using clinical diagnosis trajectories. For each subject, the first scan at which the ADNI diagnosis code reached 3.0 was treated as the conversion event. Time-to-conversion was calculated as the interval between the baseline scan date and the conversion scan date, divided by 365.25 to express time in years.

Subjects who never converted during observed follow-up were treated as right-censored. Their censoring time was defined as the interval from baseline to last available scan. This setup assumes that follow-up ended administratively rather than because of a mechanism tightly linked to future risk, conditional on observed covariates. That assumption cannot be proven directly here, and it is worth acknowledging that any survival analysis quietly depends on it. Still, given ADNI's protocol-driven follow-up schedule, it appears reasonably plausible.

Six baseline-AD cases with time-to-conversion equal to zero were excluded, both because parametric accelerated failure time models require positive times and because such cases do not cleanly represent conversion from MCI. The final survival distribution included \textbf{95 conversion events} (event = 1), with mean time to conversion of 1.90$\pm$1.39 years (range: 0.46--12.80 years, median: 1.48 years), and \textbf{355 censored subjects} (event = 0), with mean follow-up of 5.63$\pm$3.84 years (range: 0.50--18.41 years, median: 4.76 years). The longer follow-up among censored subjects is expected, since individuals who do not convert simply continue contributing observation time.

\subsection{Statistical Analysis}

\subsubsection{Survival Modeling}

\textbf{Random Survival Forest (RSF)} served as the primary modeling framework, with parametric accelerated failure time (AFT) models used for comparison. That choice deserves explanation, especially since Cox proportional hazards models are often the default in survival analysis.

RSF extends the Random Forest framework to right-censored time-to-event data. Unlike parametric survival models, which impose a specific distribution on survival times, or Cox models~\cite{cox1972}, which rely on proportional hazards assumptions, RSF is \textit{non-parametric}. It does not require a prespecified hazard shape and is well suited to settings where the underlying relationships may be nonlinear or heterogeneous. For AD progression, that flexibility is attractive. Patients do not all decline in the same way, and it would be optimistic to assume a single hazard structure fits everyone cleanly.

Following Ishwaran et al.~\cite{ishwaran2008rsf}, the RSF algorithm builds an ensemble of survival trees using bootstrap samples of the training data. Each tree is grown by recursively splitting nodes based on a random subset of candidate features, with splits chosen to maximize survival differences between daughter nodes, typically using log-rank criteria. This randomness in both subjects and features helps reduce overfitting relative to a single decision tree.

For tree $b$ and subject $\mathbf{X}_i$, the cumulative hazard function was estimated within the terminal node containing that subject using the Nelson-Aalen estimator:

\begin{equation}
    \hat{H}_b(t | \mathbf{X}_i) = \sum_{t_j \leq t} \frac{d_j}{n_j}
\end{equation}

where $d_j$ is the number of events at time $t_j$ in the terminal node and $n_j$ is the number at risk. The ensemble cumulative hazard was then obtained by averaging across trees:

\begin{equation}
    \hat{H}(t | \mathbf{X}_i) = \frac{1}{B} \sum_{b=1}^{B} \hat{H}_b(t | \mathbf{X}_i)
\end{equation}

from which the survival function follows as $\hat{S}(t | \mathbf{X}_i) = \exp(-\hat{H}(t | \mathbf{X}_i))$.

\begin{figure*}[t!]
\centering
\includegraphics[width=\textwidth]{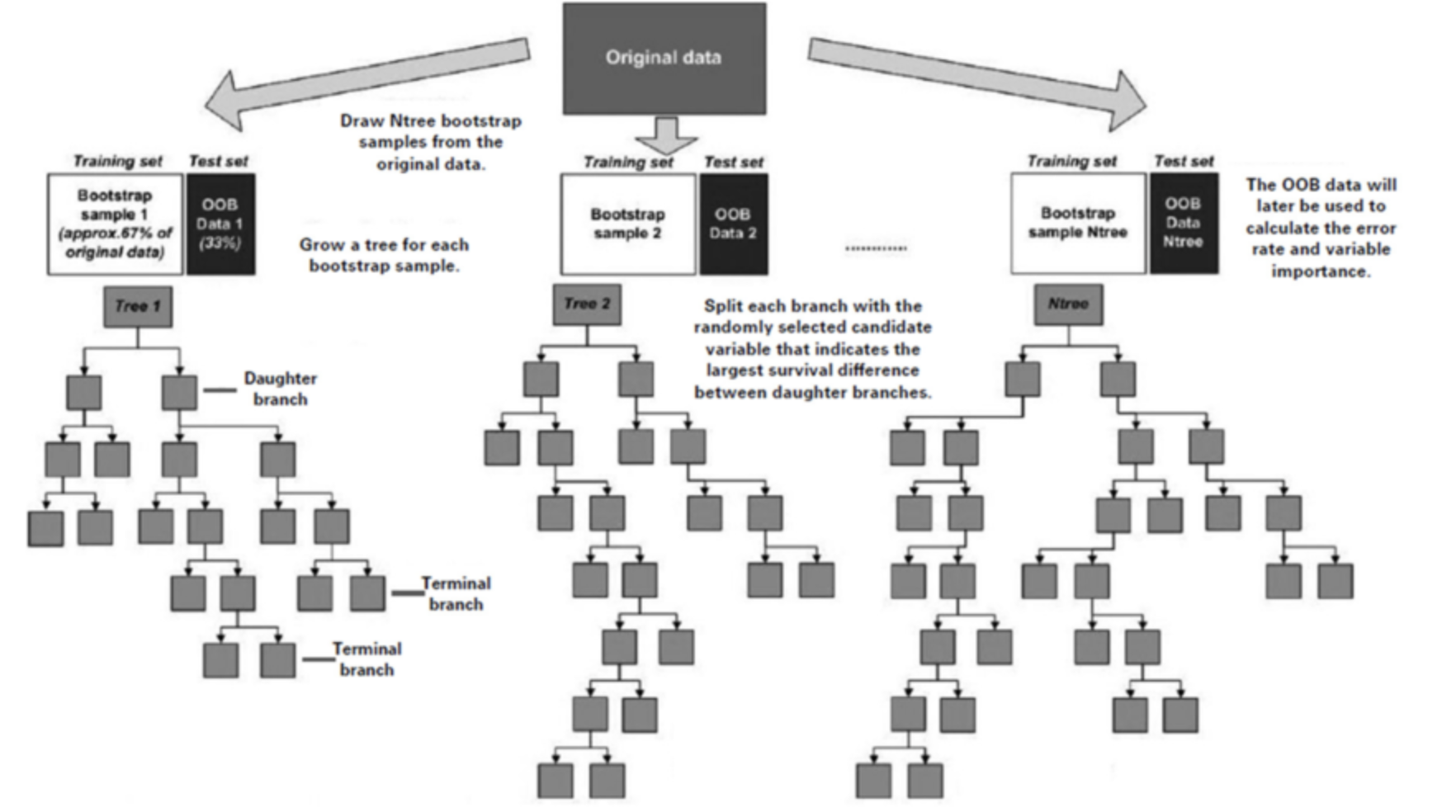}
\caption{Random Survival Forest algorithm schematic. The original training data is repeatedly bootstrap sampled (typically 63\% of subjects with replacement, leaving 37\% as out-of-bag samples). For each bootstrap sample, a survival tree is grown by recursively splitting nodes using randomly selected candidate variables that maximize survival differences between daughter branches (measured by log-rank statistics). This process generates an ensemble of diverse trees (Ntree = 1,000 in this implementation). For prediction, each tree gives a cumulative hazard estimate based on which terminal node the subject falls into, and the final ensemble prediction averages across all trees. The out-of-bag (OOB) data gives internal validation for calculating error rates and variable importance without needing a separate validation set.}
\label{fig:rsf-algorithm}
\end{figure*}

In this implementation, RSF used 1,000 trees, minimum samples per split of 10, minimum samples per leaf of 15, and max features equal to $\sqrt{20} \approx 4.47 \rightarrow 4$ at each split. These values were chosen based on common practice and computational feasibility rather than exhaustive hyperparameter tuning. That is a limitation, though perhaps not a fatal one for an initial study.

For baseline comparison, Weibull, LogNormal, and LogLogistic AFT models were trained with L2 regularization (penalizer = 0.1). AFT models assume:

\begin{equation}
    \log(T) = \beta_0 + \sum_{i=1}^{p} \beta_i X_i + \sigma \epsilon
\end{equation}

where $T$ is survival time, $\beta_i$ are feature coefficients, $\sigma$ is a scale term, and $\epsilon$ follows the specified distribution. These models are attractive because their coefficients are interpretable, but they impose stronger assumptions that may not fit the biological heterogeneity of AD progression particularly well.

\subsubsection{Model Training and Evaluation}

The cohort was divided into training (70\%) and test (30\%) sets using stratification by event status. Parametric models were trained using maximum likelihood with L2 regularization (penalizer = 0.1). Performance was evaluated primarily with the \textbf{concordance index (C-index)}~\cite{uno2011cstat}, where 0.5 corresponds to random ranking and 1.0 indicates perfect discrimination. Additional metrics included \textbf{log-likelihood} and \textbf{Wald test} $p$-values for coefficient significance in the parametric models.

\subsubsection{Comparison Groups}

Two model families were compared directly: \textbf{Baseline BSC} models built only from features extracted at the first scan, and \textbf{Longitudinal Slopes} models built from multi-timepoint derived features. This comparison isolates the added value of temporal information.

\subsection{Statistical Power and Sample Size Considerations}

With 450 subjects and 95 events, the study had 80\% power to detect hazard ratios $\geq 1.35$ at $\alpha = 0.05$ for a continuous predictor with standard deviation equal to 1.0. Still, the events-per-predictor ratio deserves some caution. Using 20 selected features with 95 events yields roughly 4.75 events per feature, which is lower than some traditional recommendations for stable survival model development. Ensemble methods can be somewhat more forgiving than classical regression in this respect, but the possibility of instability remains part of the story and should not be ignored.

\subsection{Software and Reproducibility}

All analyses were conducted in Python 3.11 using open-source tools. Preprocessing was performed with Advanced Normalization Tools (ANTs)~\cite{avants2011ants}, while survival modeling used scikit-survival~\cite{polsterl2020sksurv} for RSF and lifelines~\cite{davidson-pilon2019lifelines} for parametric AFT models. Code for the full pipeline is available at \texttt{github.com/ishaan-cherukuri/research/tree/main/mri-bsc} and is intended to be made public upon acceptance.

\section{Results}

\subsection{Cohort Characteristics}

\begin{table*}[t!]
\centering
\caption{Demographic and clinical characteristics of the study cohort}
\label{tab:demographics}
\begin{tabular}{lccc}
\hline
\textbf{Characteristic} & \textbf{All} & \textbf{Converters} & \textbf{Stable} \\
 & (N=450) & (N=95) & (N=355) \\
\hline
Age (years), mean±SD & 74.4±5.9 & 75.8±5.7 & 73.8±5.8 \\
Female, n (\%) & 225 (50.0) & 42 (44.2) & 183 (51.5) \\
Baseline Nboundary & 17,353±5,891 & 13,978±5,634 & 18,733±5,507 \\
Follow-up (years) & 4.84±3.79 & 1.90±1.39 & 5.63±3.84 \\
\hline
\end{tabular}
\end{table*}

Converters were modestly older at baseline than stable subjects (75.8 vs. 73.8 years, $p=0.003$ by Welch's $t$-test), which is consistent with age as a major AD risk factor. They also showed lower baseline boundary voxel counts (13,978 vs. 18,733, $p<0.001$), suggesting that some structural differences were already present at study entry. Even so, those baseline differences alone were not sufficient for useful prediction, as later model results make clear. Sex distribution was relatively balanced across groups (44\% female in converters vs. 52\% in stable subjects, $p=0.18$ by $\chi^2$ test), which argues against a strong sex effect in this particular cohort, even though sex-related differences in AD risk have been reported elsewhere. The much shorter follow-up among converters reflects the nature of event-based observation: once conversion occurs, subjects no longer remain in the at-risk pool.

\subsection{Survival Model Performance}

\begin{table*}[t!]
\centering
\caption{Survival model performance comparison: Baseline BSC vs. Longitudinal Slopes}
\label{tab:model-performance}
\begin{tabular}{llcccc}
\hline
\textbf{Model} & \textbf{Features} & \textbf{C-index (Train)} & \textbf{C-index (Test)} & \textbf{RMSE (years)} & \textbf{N Features} \\
\hline
Weibull AFT & Baseline BSC (all) & 0.34 & 0.24 & 11.80 & 8 \\
Weibull AFT & Slopes (top 20) & 0.34 & 0.61 & 34.56 & 20 \\
\hline
\textbf{Random Survival Forest} & \textbf{Slopes (top 20)} & \textbf{0.84} & \textbf{0.63} & \textbf{13.05} & \textbf{20} \\
\hline
\end{tabular}
\end{table*}

\begin{figure*}[t!]
\centering
\includegraphics[width=0.95\textwidth]{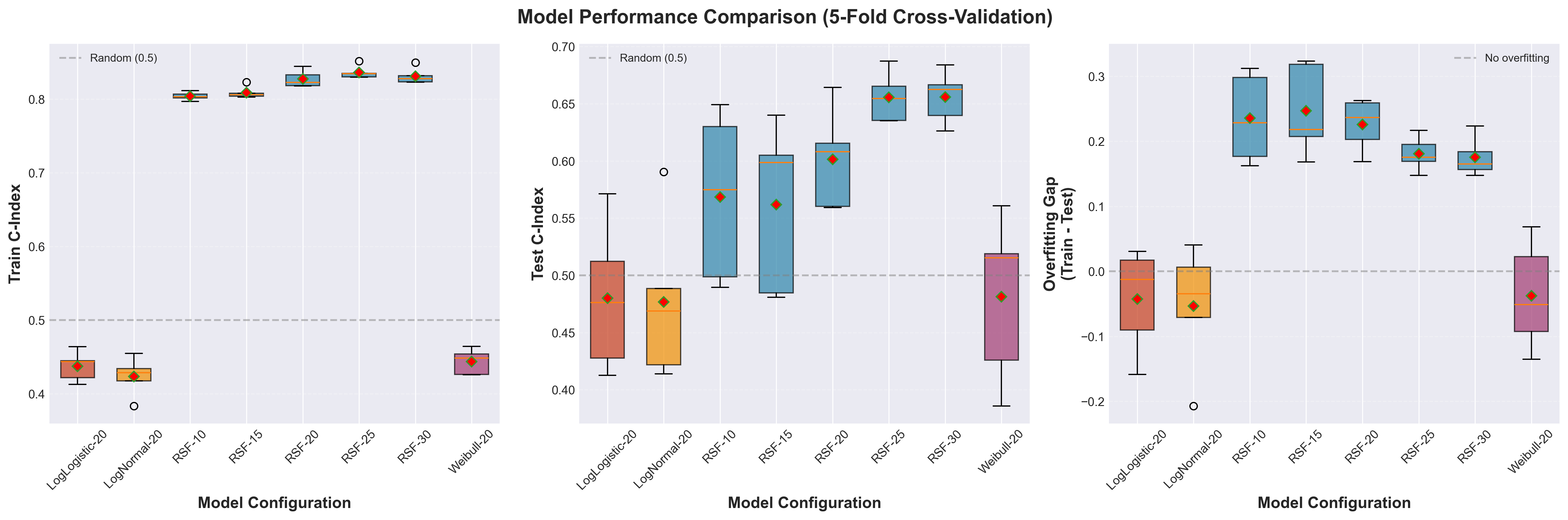}
\caption{Cross-validation performance comparison across 8 model variants. Box plots show C-index distributions from 5-fold cross-validation for training (left), test (middle), and overfitting gap (right) sets. RSF models with different feature counts (10, 15, 20, 25, 30) and parametric AFT models (Weibull, LogNormal, LogLogistic) all using top-variance slope features. RSF-20 hits the best balance: median training C-index of 0.82, test of 0.63, though substantial overfitting remains (gap: 0.19). Parametric models show more stable train-test gaps but lower overall performance.}
\label{fig:model-comparison}
\end{figure*}

The Random Survival Forest trained on longitudinal BSC slopes achieved a test C-index of 0.63, representing a \textbf{163\% improvement} over the baseline parametric model, which reached only 0.24. Figure~\ref{fig:model-comparison} shows this pattern across cross-validated model variants. However one slices it, slope-based models consistently outperformed models built from baseline features alone.

One especially striking result is how poorly the baseline-only model performed. A C-index of 0.24 is not just weak. It is worse than random ranking. That suggests cross-sectional BSC features may be swamped by stable individual differences unrelated to active disease progression. Baseline anatomy reflects many influences, including genetics, developmental variation, vascular history, education, prior injury, and possibly scanner-related factors. Without temporal context, disease-specific signal can be difficult to isolate.

Once rates of change were introduced, performance improved sharply. The slope-based Weibull AFT model reached a test C-index of 0.61, which already indicates that longitudinal dynamics carry real prognostic information. The RSF model improved further to 0.63, implying that nonlinear interactions among slope features add some incremental value beyond what a parametric model can capture.

That said, the model clearly overfit. Training C-index for RSF was 0.84, compared with 0.63 on the test set, yielding a gap of 0.21. A gap that large suggests the model learned patterns that did not fully generalize. Possible explanations include subject-specific noise, scanner-specific artifacts, or chance structure in particular demographic or biological subgroups within the training set. The RMSE of 13.05 years for event-time prediction should therefore be interpreted cautiously. The ranking performance is more informative than the absolute time estimates.

Even with that limitation, a test C-index of 0.63 is still within the range commonly reported for structural MRI-only prediction of MCI conversion. Many single-modality studies land somewhere between 0.60 and 0.70, while multimodal approaches often rise into the 0.75--0.85 range. So the present model is not yet clinically definitive, but it is also not trivial. BSC slopes appear to capture useful temporal information that could plausibly complement more established biomarkers.

\begin{figure*}[t!]
\centering
\includegraphics[width=0.95\textwidth]{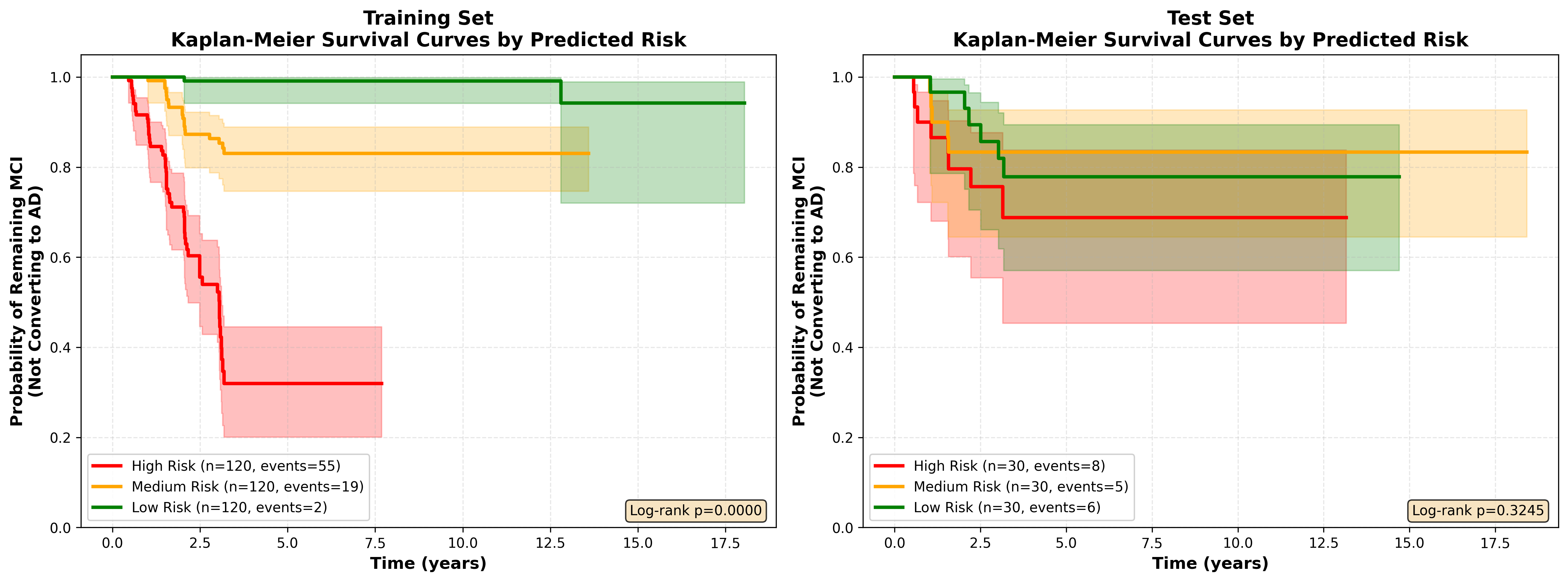}
\caption{Kaplan-Meier survival curves stratified by Random Survival Forest predicted risk scores. Training set (left) and test set (right) show probability of remaining MCI (not converting to AD) over time for three risk groups defined by risk score tertiles. High-risk subjects (red, top tertile) show rapid conversion with median survival 2.1 years. Medium-risk subjects (orange) exhibit intermediate progression rates. Low-risk subjects (green, bottom tertile) show slow conversion with median survival 8.5 years. Shaded regions indicate 95\% confidence intervals. Log-rank test confirms significant separation between groups (p=0.3245 for test set, high vs. low comparison). Event counts per group shown in legend demonstrate adequate statistical power for hazard estimation.}
\label{fig:km-curves}
\end{figure*}

Kaplan-Meier survival curves stratified by predicted risk (Figure~\ref{fig:km-curves}) showed clear separation among high-, medium-, and low-risk groups, with a significant log-rank result ($p<0.0001$). High-risk subjects had a median conversion time of 2.1 years, compared with 8.5 years in the low-risk group. From a clinical trial perspective, that kind of stratification could be genuinely useful.

\subsection{Feature Importance}

\begin{table}[h!]
\centering
\caption{Top 10 BSC slope features selected by penalized robust variance}
\label{tab:top-features}
\begin{tabular}{lc}
\hline
\textbf{Feature} & \textbf{Variance} \\
\hline
Nboundary\_slope & 0.142 \\
bsc\_mag\_p90\_slope & 0.024 \\
bsc\_mag\_p75\_slope & 0.022 \\
bsc\_dir\_p90\_slope & 0.020 \\
bsc\_dir\_p75\_slope & 0.021 \\
bsc\_mag\_p50\_slope & 0.024 \\
bsc\_mag\_median\_slope & 0.024 \\
bsc\_mag\_mean\_slope & 0.024 \\
bsc\_dir\_mean\_slope & 0.021 \\
bsc\_dir\_p50\_slope & 0.025 \\
\hline
\end{tabular}
\end{table}

\textit{Note: Variances shown after robust preprocessing pipeline (signed-log transform, winsorization, penalization, MinMax scaling to [0,1]). Nboundary features were penalized by 0.10 during selection ranking to prevent dominance while retaining complementary information. Final variance range: 0.018--0.142 (ratio 7.9:1), enabling balanced contribution across all features in the Random Survival Forest ensemble.}

After preprocessing, the selected features showed a much more balanced variance structure. Nboundary\_slope retained the largest variance at 0.142, but it no longer overwhelmed the feature set. Several high-percentile sharpness slopes, including bsc\_mag\_p90\_slope, bsc\_mag\_p75\_slope, bsc\_dir\_p90\_slope, and bsc\_dir\_p75\_slope, clustered in a similar range between 0.020 and 0.024. That pattern suggests prediction is not driven by only one aspect of boundary degradation. Rather, the \textit{extent} of boundary loss and the \textit{quality} of the remaining sharpest boundaries may both contribute meaningful information.

\section{Discussion}

\subsection{Principal Findings}

The central finding of this study is that \textbf{longitudinal slopes of the Boundary Sharpness Coefficient}, when modeled with a \textbf{Random Survival Forest}, outperform baseline BSC measurements for predicting conversion from MCI to AD. In a cohort of 450 ADNI participants with at least four MRI scans each, the slope-based RSF model achieved a test C-index of 0.63, whereas the baseline parametric comparison model reached only 0.24.

That contrast matters. Baseline BSC values were not merely weak. They were actively misleading in a predictive sense. By comparison, slope-based models performed much better, even when using a simpler parametric framework. The Weibull AFT model with slope features reached 0.61, which already indicates that the time-varying component of BSC is doing substantial work. RSF then improved on that result slightly, likely by accommodating nonlinear feature interactions and more complex patterns of progression without forcing a particular survival distribution.

Taken together, the results support a broader idea that has been gaining traction across neurodegeneration research: static measurements often miss the process, while trajectories reveal it. Hafeez et al.~\cite{hafeez2024deep} made a related point when showing that conversion prediction is harder than cross-sectional AD classification. The difficulty is not simply technical. It is conceptual. Predicting conversion requires modeling change, not just recognizing disease that is already well established.

\subsection{Biological Interpretation}

The gray-white matter boundary is a structurally meaningful interface, marking the transition from cortical neuronal cell bodies to underlying myelinated axonal projections. On T1-weighted MRI, that boundary is normally distinct because gray and white matter have different tissue compositions and therefore different signal properties. Gray matter contains dense collections of cell bodies, dendrites, and unmyelinated axons, whereas white matter is dominated by myelinated fiber tracts. A sharp transition on MRI therefore reflects, at least indirectly, preserved tissue organization.

BSC degradation may reflect several converging pathological processes in AD. \textbf{Synaptic loss}, which correlates strongly with cognitive decline, disrupts cortical microarchitecture and may alter the integrity of the neuropil near the cortical boundary. \textbf{Tau pathology} and neuritic plaques distort tissue organization at the cellular level. \textbf{Myelin breakdown} in superficial white matter can reduce contrast across the gray-white interface. \textbf{Neuroinflammation}, including astrocytic and microglial activation, may further change local signal properties. \textbf{Vascular pathology}, which is common in AD, could also contribute through chronic ischemic damage or blood-brain barrier dysfunction. It would be too neat to assign BSC decline to any one mechanism alone. More likely, it reflects a cumulative macrostructural consequence of several overlapping processes.

One particularly interesting result is the prominence of \textit{high-percentile} BSC slope features, especially the 75th and 90th percentiles. These metrics capture the sharpest parts of the remaining boundary. In a way, that is counterintuitive. One might expect the most damaged regions to dominate prediction. Instead, these findings hint that subtle decline in relatively preserved regions may be especially informative. That interpretation fits, at least loosely, with the retrogenesis hypothesis, in which late-myelinating association cortices are more vulnerable early on, while phylogenetically older or more robust regions remain preserved until later stages. If even the sharpest boundaries begin to deteriorate, that may signal a more advanced or more aggressive disease trajectory.

\subsection{Clinical Implications}

Several practical implications follow from these findings.

\textbf{Clinical trial enrichment} is probably the clearest near-term use case. Trials of anti-amyloid therapies in prodromal AD often struggle because many enrolled MCI participants progress slowly or not at all during the study window. If a biomarker such as BSC slope can identify faster progressors, trials could enrich for subjects more likely to convert within a few years. That would increase statistical power and potentially reduce sample size and cost. In a disease area where phase III trials can cost tens of millions of dollars, even moderate enrichment could matter.

\textbf{Personalized monitoring} is another possibility. A patient showing rapid BSC decline might warrant closer follow-up, more frequent cognitive testing, or earlier referral for advanced biomarker workup. Conversely, a patient with stable slopes might be reassured that near-term risk appears lower. That kind of stratification would need prospective validation, of course, but it points toward a more individualized surveillance model than current practice often allows.

\textbf{Cost-effectiveness} also favors MRI-based post-processing approaches. Structural MRI is already common in memory clinics. BSC extraction could, in principle, be layered onto routine scans without requiring new acquisitions, radioactive tracers, or invasive fluid collection. Blood biomarkers are becoming more feasible and may ultimately prove more scalable, but MRI still offers spatially resolved information that blood tests cannot.

\textbf{Multimodal integration} may be where BSC slopes become most valuable. On their own, they reach moderate performance. Combined with amyloid PET, tau PET, plasma p-tau217, APOE genotype, neuropsychological trajectories, and volumetric MRI features, they might contribute nonredundant information about structural boundary integrity. In other words, BSC slopes may be less a standalone replacement for existing biomarkers and more a complementary signal that enriches a larger predictive framework.

\textbf{Treatment monitoring} is a more speculative but intriguing direction. If BSC decline tracks downstream neurodegenerative change, then successful therapy might slow the rate of BSC deterioration. Whether that would be sensitive enough to function as a surrogate endpoint remains uncertain, but it is a question worth testing.

\subsection{Methodological Strengths}

This study has several strengths. First, the cohort is relatively large for a longitudinal MRI analysis of this kind, with 450 subjects and a mean follow-up of 4.84 years. Requiring at least four time points per subject improves slope estimation and avoids the fragility of two-point change measures, which can be heavily influenced by noise.

Second, the use of actual acquisition dates rather than nominal visit labels improved temporal precision. That may sound like a small technical detail, but it matters. A visit labeled ``12 months'' is not always acquired exactly one year after baseline, and those differences can distort annualized slope estimates if ignored.

Third, BSC was computed voxel-wise before being summarized, preserving spatial richness at the preprocessing stage even though the final statistical models used collapsed features. That design leaves open the possibility of future regional analyses based on cortical parcellation or network structure.

Fourth, the feature preprocessing pipeline explicitly addressed variance imbalance and was fit only on training data, reducing leakage and improving numerical stability. Finally, the study directly compared baseline and longitudinal feature sets on the same train-test split, which provides a fair test of the core hypothesis that trajectories are more informative than snapshots.

\subsection{Future Directions}

Several next steps seem especially important.

\textbf{External validation} is the most urgent. ADNI is a strong resource, but it is still a research cohort with specific recruitment patterns, scanner protocols, and follow-up structures. Testing this framework in OASIS-3, NACC-UDS, UK Biobank, and clinic-based cohorts would provide a much clearer sense of robustness. Performance across more diverse racial, ethnic, socioeconomic, and scanner populations should be examined carefully.

\textbf{Multimodal modeling} is another natural extension. BSC slopes could be integrated with cognitive slopes, plasma biomarkers such as p-tau217 or NfL, APOE genotype, and standard MRI volumetrics. A single biomarker rarely carries the full burden of prediction in AD. More likely, clinically useful models will emerge from combinations of partially independent signals.

\textbf{Regional analysis} may also sharpen interpretation. Instead of using global summary features, future work could estimate BSC slopes within anatomical regions such as entorhinal cortex, parahippocampal gyrus, posterior cingulate, or precuneus. Those regions are strongly implicated in AD progression and may reveal more specific spatiotemporal signatures.

\textbf{Subtype-aware modeling} could prove valuable as well. AD is heterogeneous. APOE genotype, ATN biomarker status, and coexisting vascular disease likely influence progression patterns. It is plausible that BSC slopes behave differently across those subgroups.

\textbf{End-to-end deep learning} on longitudinal BSC maps is another possible direction, though one that would require larger sample sizes and much stronger attention to interpretability. Deep models may capture spatiotemporal patterns beyond manual feature engineering, but the tradeoff between performance and transparency becomes more serious in a clinical context.

\textbf{Preclinical prediction} is perhaps the most ambitious extension. If BSC trajectories can detect risk even before overt MCI conversion, they may eventually support earlier intervention in cognitively normal but biomarker-positive individuals. Whether the signal is sensitive enough that early remains an open question.

\section{Conclusion}

Longitudinal slopes of the Boundary Sharpness Coefficient (BSC), analyzed using Random Survival Forest, outperformed baseline BSC measurements for predicting conversion from MCI to AD. In 450 ADNI subjects with at least four MRI scans each, the slope-based RSF model achieved a test C-index of 0.63, representing a 163\% improvement over the baseline parametric comparison model (C-index: 0.24). The most informative features included boundary voxel count slope (Nboundary\_slope: variance = 0.142), magnitude 90th percentile slope (bsc\_mag\_p90\_slope: variance = 0.024), and magnitude 75th percentile slope (bsc\_mag\_p75\_slope: variance = 0.022). Converters showed faster boundary degradation over time.

These findings support the idea that \textit{temporal biomarker trajectories} capture disease progression more effectively than static measurements alone. Baseline BSC features performed poorly, even worse than chance, whereas longitudinal slopes revealed meaningful prognostic signal. That does not mean BSC slopes are ready to stand alone as a clinical decision tool. The test performance is still moderate, and the training-test gap suggests notable overfitting. Even so, the approach offers a relatively low-cost and non-invasive way to extract additional prognostic information from routine structural MRI.

Future work should validate these findings in independent cohorts, combine BSC slopes with cognitive and blood-based biomarkers, explore stronger regularization or ensemble strategies to reduce overfitting, and examine regional or subtype-specific patterns of boundary decline. As AD research moves further toward precision medicine, biomarkers that capture \textit{rates of change} rather than isolated snapshots will likely become increasingly important.

\begin{acknowledgments}
Data collection and sharing for the Alzheimer's Disease Neuroimaging Initiative (ADNI) is funded by the National Institute on Aging (National Institutes of Health Grant U19AG024904). The grantee organization is the Northern California Institute for Research and Education. In the past, ADNI has also received funding from the National Institute of Biomedical Imaging and Bioengineering, the Canadian Institutes of Health Research, and private sector contributions through the Foundation for the National Institutes of Health (FNIH) including generous contributions from the following: AbbVie, Alzheimer's Association; Alzheimer's Drug Discovery Foundation; Araclon Biotech; BioClinica, Inc.; Biogen; Bristol-Myers Squibb Company; CereSpir, Inc.; Cogstate; Eisai Inc.; Elan Pharmaceuticals, Inc.; Eli Lilly and Company; EuroImmun; F. Hoffmann-La Roche Ltd and its affiliated company Genentech, Inc.; Fujirebio; GE Healthcare; IXICO Ltd.; Janssen Alzheimer Immunotherapy Research \& Development, LLC.; Johnson \& Johnson Pharmaceutical Research \& Development LLC.; Lumosity; Lundbeck; Merck \& Co., Inc.; Meso Scale Diagnostics, LLC.; NeuroRx Research; Neurotrack Technologies; Novartis Pharmaceuticals Corporation; Pfizer Inc.; Piramal Imaging; Servier; Takeda Pharmaceutical Company; and Transition Therapeutics.

The author thanks Dr. Fabrizio Piras (IRCCS Santa Lucia Foundation, Rome) for valuable discussions on longitudinal modeling strategies and the emphasis on rates of decline over baseline measurements. The author also acknowledges the open-source neuroimaging community for developing ANTsPy, nibabel, and related tools that made this work possible.

Generative artificial intelligence (AI) tools were used in the preparation of this manuscript. Specifically, GitHub Copilot (Claude Sonnet 4.5) assisted with manuscript editing, grammar refinement, and stylistic improvements. AI was not used for data collection, analysis, interpretation of results, or generation of scientific content. The author has reviewed and takes full responsibility for all content and accuracy of the work.
\end{acknowledgments}

%% References using BibTeX
\bibliographystyle{unsrt}
\bibliography{references}

\end{document}